\documentclass[pra,floatfix,twocolumn]{revtex4}
\usepackage[dvips]{graphicx}%
\usepackage{amsmath}
\usepackage{bm,color} 
\usepackage{amsmath,amssymb}
\newcommand{\braket}[2]{\langle #1 | #2 \rangle}
\newcommand{\ket}[1]{\left |  #1 \right \rangle}

\newcommand{\ave}[1]{ \langle #1   \rangle}
\def \tr{{\textrm {Tr}}}

\begin{document}
\title{Photonic  families of non-Gaussian entangled states and entanglement criteria for continuous-variable systems}

\author{Ryo Namiki}
\affiliation{Department of Physics, Graduate School of Science, Kyoto University, Kyoto 606-8502, Japan}
 \date{May 30, 2012}
\begin{abstract} 
We consider two classes of non-Gaussian entangled states generated from the product of number states with the action of the beamsplitter or the two-mode squeezer. It is shown that,  for many of these states,  the covariance matrix is compatible with the covariance matrix of separable Gaussian states and their separability cannot be verified by the measurements of the first and second moments of canonical variables. We identify a couple of continuous-variable entanglement criteria with higher order moments to verify these non-Gaussian entanglement. 
\end{abstract}

\maketitle


\section{Introduction}
  
  Gaussian states, such as coherent states, squeezed states, and two-mode squeezed states, are  of fundamental to describe basic nature of quantum mechanical phenomena. They play a central role in experiments of quantum optics and quantum information science \cite{NC00,B2000,book1,RMP-CV,AdessoRev}. Gaussian states are characterized by their covariance matrices with the first and second order moments of canonical quadrature variables. Similarly, Gaussian operations are    characterized by the transformation of the covariance matrices, and    there have been various experimental demonstrations of quantum information processing within the Gaussian framework \cite{book1,RMP82}.
In turn, it has been known that  non-Gaussian resource is necessary to implement a universal quantum computation \cite{Lloyd99,Meni06,Gu09}. This fact strongly motivates us to study the property of non-Gaussian quantum states.

 There have been many works to investigate the properties of non-Gaussian wave functions \cite{TMSN,NBS} and non-Gaussian entangled states \cite{AdessoRev,Kitagawa06,Dell06,Adesso09,Lee10,Lee11,NGsep,namiki10,Kim02,Go09}. 
 An outstanding example of non-Gaussian quantum states is a single photon state. Generation method of the single photon state and its non-Gaussian properties have been widely investigated not only for its primitive quantum nature 
but also for potential application in quantum information science \cite{R09}.  
 Quantum measurements  to manipulate a few number of photons  have been experimentally demonstrated \cite{Gran04,Zava08,Takahashi08,Name10} and such methods are useful to generate non-Gaussian entanglement \cite{Kitagawa06,Lee10,Lee11,Takahashi10,Gran07}.

 Entanglement of Gaussian states has been successfully described by their covariance matrices \cite{Gie03,RMP-CV,AdessoRev,Duan-Simon,Simon}.   On the other hand, there have been many proposed entanglement criteria based on the measurements of higher order moments of canonical quadrature variables \cite{SV1,Aga05,SV2,Sun11,HZ06,SV3,SV4,Nha07}.   
 Generally, the criteria with higher order moments are thought to be powerful to detect non-Gaussian entanglement, however, the criteria based on the covariance matrices are sensitive to some of non-Gaussian entangled states. Thus, it is not so clear in what condition we need  higher order moments for entanglement verification.   Hence, it would be valuable to investigate limitations of entanglement criterion with the lower order moments of the covariance matrix and show the advantage of higher order criteria \cite{Dell06,Go09}. Thereby, the following two steps would be important:  
   (i) To identify a set of non-Gaussian entangled states whose inseparability  cannot be verified by the measurement of the covariance matrix. 
   (ii) To find  higher order entanglement criterion which can verify the inseparability of these states. 
      
    The photonic number states are eigenstates of a harmonic oscillator and their properties are simply described by using the annihilation and creation operators. Such a formalism enables us to describe  an orthonormal basis consist of entangled states \cite{namiki10}. It is likely that such an orthonormal basis gives an insightful example to proceed the step (i). To proceed  the step (ii), there have been a sequence of works to present the separable conditions based on SU(2) and SU(1,1) commutation relations \cite{HZ06,Aga05,Nha07,Sun11,SV3}. These conditions can be derived from the uncertainty relations for the operators described by the products of canonical variables, and are thought to be the lowest order criteria to go beyond the Gaussian framework.

In this paper, we consider two classes of non-Gaussian entangled states generated from the product of the number states by using two-mode Gaussian unitary interactions.
 We show that, for many of these states, the covariance matrix is compatible with the covariance matrix of separable Gaussian states and their inseparability cannot be verified by the measurements of the first and second moments of canonical variables. We also identify a couple of continuous-variable entanglement criteria with higher order 
  moments to verify the inseparability of these states. 
 
This paper is organized as follows. We investigate the property of the covariance matrix for the  non-Gaussian entangled states in Sec. II. We consider the entanglement criteria with higher order moments for the  non-Gaussian entangled states in Sec. III. The results are summarized in Sec IV.

\section{Photonic non-Gaussian entangled states and entanglement criteria for Gaussian states}
In this section we apply the separable condition based on the covariance matrix of the density operator to two families of non-Gaussian entangled states, and show limitations on the detection of entanglement from the measurement of the covariance matrix.
\subsection{Covariance matrix and inseparability of two-mode states}
We consider a two-mode system $AB$ described by quadrature variables with the canonical commutation relations  $[\hat x_A, \hat  p_A]=[ \hat x_B, \hat  p_B]=i$. 
Let us define a vector form for the set of  quadrature variables as
\begin{eqnarray}
\hat R := (\hat x_A, \hat p_A, \hat x_B, \hat p_B )^t.  
\end{eqnarray}
The covariance matrix of a two-mode density operator $ \rho$ is defined by
\begin{eqnarray}
\gamma:&=& \langle \hat R \hat R^t+(\hat R \hat R^t)^t \rangle -2\langle \hat R \rangle \langle  \hat R^t \rangle \nonumber \\
&=& \langle (\Delta \hat R )(  \Delta  \hat R^t) +[( \Delta  \hat R ) ( \Delta  \hat R^t)]^t \rangle  = : \left(
  \begin{array}{cc}
    A   & C   \\
     C^t  & B   \\
  \end{array}
\right),  \label{covama}
\end{eqnarray}
where  $\ave{ \hat O } =\tr (\hat O \rho )$ denotes   the expectation value of the observable $\hat O$ and $ \Delta \hat  O := \hat  O - \ave{\hat O}$. The submatrices $A$, $B$, and $C$ are real symmetric 2-by-2 matrices. Their determinants and $\det \gamma$ are invariant under the local Gaussian unitary transformations. 
In terms of these determinants, the separable criterion \cite{Simon} is given by  
\begin{eqnarray}
D:= \det \gamma +1 -\det A -\det B +2 \det C \ge 0. \label{sep1}
\end{eqnarray} 
This is the necessary and sufficient condition for separability of two-mode Gaussian states. For non-Gaussian states, this is a necessary condition for separability. 
Hence, we may define a class of purely non-Gaussian entangled states by the set of entangled states which fulfill the condition of Eq. (\ref{sep1}).  
An important property of this class of states  is that their covariance matrices cannot be transform into the covariant matrices of the entangled Gaussian states by local Gaussian operations and classical communications.

\begin{figure}[htb]
  \includegraphics  [width=0.9\linewidth]{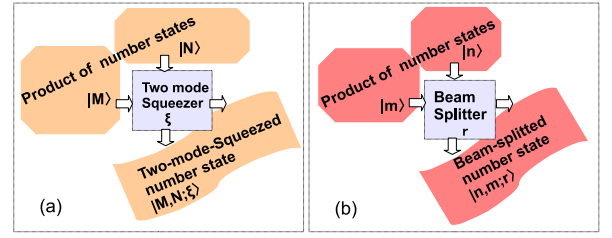}
  \caption{Two families of photonic non-Gaussian entangled states from a product of number states.  (a) The two-mode squeezed number states are generated by the action of the two-mode squeezing operation. (b) The beamsplitted number states are generated by the beam-splitter operation. The parameters $\xi$ and $r $ specify the action of the two-mode squeezer and the action of the beamsplitter, respectively (see main text).  }%
    \label{fig:sep-fig1.eps}
\end{figure}

In the next two subsections we consider two families of entangled states given by  acting  Gaussian unitary interactions on the product of number states as in FIG. \ref{fig:sep-fig1.eps}. We will specify the regime of parameters which indicates  the purely non-Gaussian entangled states in each family.
 
 \subsection{Two-mode-squeezed number states}
Let us write the annihilation operators of mode $A$ and mode $B$ by $a  = (\hat x_A + i\hat p_A)/\sqrt 2 $ and $b  = (\hat x_B + i\hat p_B)/\sqrt 2 $. This implies the bosonic commutation relations $[a,a^\dagger ]=1$, $[b,b^\dagger ]=1$, and $[a,b ]= [a,b^\dagger ]=0 $. The number states of the local modes are defined by  $ | n \rangle_{A } :=   (a^\dagger )^n|0 \rangle_A /{\sqrt{n!}} $ and $| n \rangle_{B } :=  {(b^\dagger )^n}|0 \rangle_B/{\sqrt{n!}}  $ together with the conditions for the vacuums, $a \ket{0 }_A=0$ and $b \ket{0 }_B=0$.  By  using this standard notation, 
the two-mode-squeezed (TMS) vacuum state  in the Schmidt decomposed form is defined by
\begin{eqnarray}
| \psi_\xi \rangle_{AB}  :=  \sqrt{1-|\xi| ^2} \sum_{n= 0}^\infty\xi ^n    |n \rangle  _A |  n \rangle_B, \label{defTMSV}  \end{eqnarray}
where we assume $ |\xi|<1  $. The TMS vacuum state is a Gaussian state and it is known that any two-mode pure Gaussian state can be transformed into the form of Eq. (\ref{defTMSV}) by  local Gaussian unitary operations.

 Let us define the annihilation operators crossing over the separation of $A$ and $B$ by 
\begin{eqnarray}
 \hat A_\xi  :=  V_\xi a  V_\xi^\dagger  =  \frac{a  -\xi b ^\dagger }{\sqrt{1-|\xi|^2}},\   
\hat B_\xi  =  V_\xi b  V_\xi^\dagger  = \frac{b  -\xi a^\dagger}{\sqrt{1-|\xi| ^2}}, \label{defA} \end{eqnarray} where the unitary operator $V_\xi:= e^{\xi a^\dagger b^\dagger - \xi^* a b}$ is the two-mode squeezing operator.  
The new field operators fulfill the bosonic commutation relations $[\hat A_\xi ,\hat A_\xi ^\dagger] =[\hat B_\xi ,\hat B_\xi ^\dagger] = 1 $ and  $[\hat A_\xi,\hat B_\xi]=[\hat A_\xi,\hat B_\xi^\dagger]= 0 $.  
The TMS vacuum state corresponds to the vacuum of the new field operators as we have $\hat A_\xi |\psi_\xi \rangle = 0$ and $\hat B_\xi |\psi_\xi\rangle =0 $.
The TMS number  state \cite{TMSN,namiki10,Dell06} is defined by the number state of the non-local modes as
\begin{eqnarray}
|M,N;\xi \rangle =\frac{ (\hat A_\xi^\dagger)^{M} }{\sqrt{M!}} \frac{(\hat B_\xi^\dagger)^{N }}{\sqrt{N !}} | \psi_\xi\rangle . \label{basisi}
\end{eqnarray} From the construction, it satisfies  the orthonormal relation $\langle M',N';\xi |M,N;\xi\rangle =\delta_{N,N'} \delta_{M,M'} $. It is entangled whenever $\xi \neq 0 $  and its wave function is non-Gaussian except for $M=N=0$ \cite{namiki10}. 
From Eqs. (\ref{defA}) and (\ref{basisi}) we can confirm the relation $|M,N;\xi \rangle =V_\xi |M \rangle_A| N\rangle_B$. This implies that the TMS number states can be generated by acting  the  two-mode squeezing operation on the product of number states as in FIG. \ref{fig:sep-fig1.eps}(a).

\begin{figure}[htb]
  \begin{center}
     \includegraphics[width=0.8\linewidth]{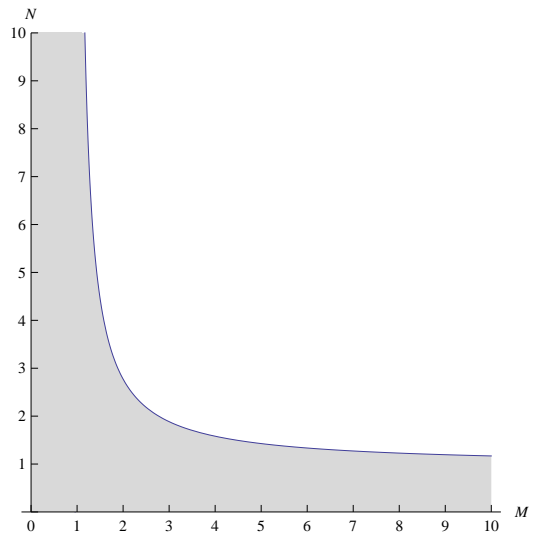}
  \end{center}
  \caption{A pair of non-negative integers ($M$, $N$) assigns a  TMS number state $\ket{M,N; \xi}$. In the gray regime, Eq. (\ref{joken1}) is fulfilled, and entanglement of the corresponding TMS number states can be verified by the measurement of the covariance matrix. On the outside (and the boundary) of this regime, the covariance matrix of $\ket{M,N;\xi} $ has to be  compatible with  a separable Gaussian state, and its inseparability cannot be confirmed by  the measurement of the covariance matrix.   Here, we set the parameter  $\xi=0.7$ in Eq. (\ref{joken1}).  } 
  \label{fig:ENS-sep-fig1.eps}
\end{figure}

The quadrature moments for the TMS number states are routinely calculated by using the following relations on the annihilation and creation operators:
\begin{eqnarray}
a&=&\frac{\hat A_\xi +\xi \hat B_\xi ^\dagger }{\sqrt{1-|\xi|^2}}, \ \ a ^\dagger=\frac{\hat A_\xi ^\dagger   +\xi^*  \hat  B_\xi}{\sqrt{1-|\xi|^2}} \nonumber\\
b&=&\frac{\hat B_\xi +\xi  \hat  A_\xi ^\dagger }{\sqrt{1-|\xi|^2}}, \ \ b  ^\dagger = \frac{\hat B_\xi ^\dagger +\xi^*   \hat  A_\xi }{\sqrt{1-|\xi|^2}}. \label{stob}
\end{eqnarray}
The elements of the covariance matrix are determined to be   
\begin{eqnarray}
A &=&\frac{ 1+2M +(1+2N )|\xi|^2 }{ 1-|\xi|^2}
 \left(
  \begin{array}{cc}
 1  &   0 \\
    0  &1   \\
  \end{array}
\right),\nonumber 
  \\
B &=&\frac{{1+2N}  + (1+2 M)|\xi|^2  }{ 1-|\xi|^2}  
 \left(
  \begin{array}{cc}
 1  &   0 \\
    0  &1   \\
  \end{array}
\right), \nonumber
 \\
C &=&\frac{2 |\xi| (1+M+N)  }{ 1-|\xi|^2}
\left(
  \begin{array}{cc}
 \cos \theta  &   \sin \theta \\
   \sin \theta   &- \cos \theta   \\
  \end{array}
\right),  
\end{eqnarray} where we write $\xi = |\xi| (\cos \theta + i \sin \theta )$.
 This implies an explicit formula of the quantity $D$ in Eq. (\ref{sep1}) as \begin{eqnarray}
D &=&\left(\frac{4}{1-|\xi|^2}\right)^2  \left( {(1+N)(1+M)|\xi|^2 -NM}\right) \nonumber \\ && \times {\left( MN |\xi|^2- (1+N)(1+M)\right)} .
\end{eqnarray}
From the condition $|\xi| < 1 $ we can see that the final term is negative, i.e.,  ${\left( MN |\xi|^2- (1+N)(1+M)\right)} < 0 $. 
Hence, the separable inequality $D \ge 0$ in Eq. (\ref{sep1}) is violated when 
\begin{eqnarray}
\left(M - \frac{|\xi|^2}{1-|\xi|^2} \right) \left(N - \frac{|\xi|^2}{1-|\xi|^2} \right)  < \frac{|\xi|^2}{(1-|\xi|^2)^2 }.   
   \label{joken1}
\end{eqnarray}
The parameter regime of this inequality can be depicted  in FIG. \ref{fig:ENS-sep-fig1.eps}. 
From Eq. (\ref{joken1}), 
 we can specify the subset of the TMS number states whose inseparability cannot be confirmed by the measurement of the covariance matrix.  We can see that 
   the majority of the TMS number states belong to the class of the purely non-Gaussian entangled states.

\subsection{Beam-splitted number states}
Let us consider the beam-splitter transformation described by
\begin{eqnarray}
 c_r  :=U_r a U_r ^\dagger =  \frac{a  -r b  }{\sqrt{1+|r|^2}},\   
d_r  :=U_r b U_r ^\dagger =   \frac{ r^*   a + b }{\sqrt{1+|r| ^2}},\nonumber \\  \label{defC} \end{eqnarray} where $U_r =e^{r a^\dagger b - r^* a b^\dagger }$ is the unitary operator of the beamsplitter and  we assume $ 0<|r| < \infty$. 
The field operators fulfill the bosonic commutation relations $[c_r ,c_r ^\dagger] =[d_r ,d_r ^\dagger] = 1 $ and  $[ c_r, d_r]=[  c_r, d_r^\dagger]= 0 $.  In the case of $r =1 $ we have  the  half-beamsplitter transformation
\begin{eqnarray}
 c  =  \frac{a  - b  }{\sqrt{2}},\   
d =   \frac{  a + b }{\sqrt{2 }}.  \end{eqnarray}  In the limit of $|r| \to 0 $, there is no interaction between the modes and  we have a trivial transformation $(c, d ) = ( a, b ) $ whereas we have another trivial transformation $(c, d ) = ( b, a ) $ in the limit of $|r| \to \infty $ up to the phase factor. 

The beam-splitter transformation produces another family of orthonormal states [see, FIG. \ref{fig:sep-fig1.eps}
(b)],
\begin{eqnarray}
|n,m;r \rangle : =   U_r  |n,m \rangle &=& 
 \frac{ (  c_r^\dagger)^{n} }{\sqrt{n!}} \frac{( d_r^\dagger)^{m}}{\sqrt{m !}} |0,0 \rangle . \label{p-basisi}
\end{eqnarray} We call this state a beam-splitted (BS) number state. In the case of $n=m=0$, the state $\ket{0,0;r}= \ket{0,0}$  is a separable Gaussian state. Except for this case the BS number states are non-Gaussian and entangled as we prove in the following.  

Let us assume  the  non-zero-photon case of  $n+m \ge 1$. 
Recall that any Gaussian states are non-orthogonal with each other.  Since, a non-zero-photon state $\ket{n,m;r}$ is orthogonal to the vacuum state from the orthonormal relation $\braket{n',m';r}{n,m;r}=\delta_{n,n'} \delta_{m,m'}$ it cannot be a Gaussian state.  
From Eqs. (\ref{defC}) and (\ref{p-basisi}), we can verify that the BS number state $|n,m;r \rangle$ has Schmidt decomposition with the Schmidt basis $\{\ket{n+m-k,k}\}_{k=0,1,2, \cdots n+m}$ and the Schmidt rank is at most $n+m+1$.  
It is direct to see that the coefficients of the edges of the Schmidt basis $\braket{n+m,0}{n,m;r}$ and $\braket{0,n+m}{n,m;r}$ is non-zero. 
This implies that one can find entanglement from the subspace spanned by $\ket{n+m,0}$ and $\ket{0,n+m}$.  
Hence, the BS number states are non-Gaussian entangled states except for the case of $n=m=0$. 
The inseparability of the BS number states can be also proven from the theorem for beam-splitter entangler \cite{66} (see also \cite{Kim02}). 
Note that a BS number state is described by a finite number of Schmidt bases, and  it is thought to be an entity of discrete-variable systems. On the contrary,  Schmidt rank of the TMS number states is infinite, and    they are thought to be entities of purely infinite-dimensional systems. 

The quadrature moments for the BS number states are readily calculated by  using the following relations on the annihilation and creation operators,
\begin{eqnarray}
a&=&\frac{  c _r +r  d_r  }{\sqrt{1+ |r|^2}}, \ \ a ^\dagger=\frac{ c_r ^\dagger   +r^* d_r ^\dagger }{\sqrt{1+ |r|^2}} \nonumber\\
b&=&\frac{  d _r - r^* c_r }{\sqrt{1+ |r|^2}}, \ \ b  ^\dagger = \frac{ d_r ^\dagger -r  c_r ^\dagger  }{\sqrt{1+|r|^2}}. 
\end{eqnarray}
The elements of the covariance matrix $\gamma$ of Eq. (\ref{covama})  are determined to be   
\begin{eqnarray}
A &=&\frac{ {1+|r| ^2}  + 2(n +|r|^2 m  )}{ 1+|r|^2}
 \left(
  \begin{array}{cc}
 1  &   0 \\
    0  &1   \\
  \end{array}
\right)\nonumber  ,
  \\
B &=&\frac{ {1+|r|^2}  + 2(|r|^2  n +m  )}{ 1+|r|^2}
  \left(
  \begin{array}{cc}
 1  &   0 \\
    0  &1   \\
  \end{array}
\right)\nonumber ,
 \\
C &=&\frac{ 2r (m-n )  }{ 1+ r^2}
\left(
  \begin{array}{cc}
 \cos \phi  & - \sin \phi \\
 \sin \phi      & \cos \phi   \\
  \end{array}
\right) ,  
\end{eqnarray} where $r = |r| ( \cos{\phi} + i \sin \phi ) $.
From these relations and  Eq. (\ref{sep1}), we have \begin{eqnarray}
 D   &=& \frac{16 }
{(1 + |r|^2)^2} (m (1 + n) + (1 + m) n |r| ^2) \nonumber \\ & & \times ((1 + m) n + m (1 + n) |r|^2)  .  
\end{eqnarray}
This implies that the separable condition  $D \ge 0$  of Eq. (\ref{sep1}) 
 is always satisfied for the BS number states.  
 (This result is also given by the fact $\det C  =  \frac{ 4 (m-n )^2 |r|^2 }{(1+|r| ^2)^2}\ge 0 $ with Lemma of Ref. \cite{Simon}, that is,  Gaussian states with $\det C \ge 0 $ are separable.). Hence, they are classified to the purely non-Gaussian entangled states.  

It might be interesting that a relation similar to Eq. (\ref{joken1})  can be derived for the BS number states when we consider a higher order entanglement condition given by  Hillery and  Zubairy  \cite{HZ06}: 
\begin{eqnarray}
 \ave{ a^\dagger  a b^\dagger b}  < |\ave{a b^\dagger }|^2 
 \label{jokenhz}. 
\end{eqnarray}
For the BS number states of Eq. (\ref{p-basisi}) we have 
\begin{eqnarray}
\ave{ a^\dagger  a b^\dagger b} &=& \left(\frac{1}{1+ |r|^2} \right)^2  \nonumber \\
 && \times \left\{(1- |r | ^2)^2 + |r|^2 [m(m-1)+ n(n-1)] \right\}, \nonumber \\
\ave{  a  b^\dagger }&=& \frac{r}{1+ |r|^2} (m-n) . \label{eq18}
\end{eqnarray}
Then, the entanglement condition of Eq. (\ref{jokenhz}) turns out to be
\begin{eqnarray}
 \left(m- \frac{|r|^2}{1+ |r|^4} \right) \left(n- \frac{|r|^2}{1+ |r|^4} \right) 
<  \left( \frac{|r|^2}{1+ |r|^4} \right)^2. \label{nm-qua}
\end{eqnarray}
This is an inverse proportional condition on the photon numbers of entangled modes for the inseparability similar to Eq. (\ref{joken1}) [See also, FIG. \ref{fig:ENS-sep-fig1.eps}].  The condition of Eq. (\ref{nm-qua}) is satisfied only for the cases of $ (m,n)=\{ (k,0), (0,k)\}$ with $k=1,2,3, \cdots $. 
This is an advantage of the measurements of higher order moments for detection of the purely non-Gaussian entanglement although it can detect the entanglement of a rather small portion of the BS number states.

Now, it has turned out that  the measurement of the covariance matrix is not useful to detect the entanglement for a significant portion of the photonic non-Gaussian entangled states. We thus consider higher order criteria in the subsequent section.

\section{Verification of Photonic non-Gaussian entanglement via the measurements of  higher order moments}
In this section we consider the separable conditions based on the operators that form SU(2) algebra and SU(1,1) algebra  \cite{SV3,HZ06,Aga05,Nha07,Sun11}. We show that   the measurements of quadrature moments to the fourth order are sufficient to verify the present non-Gaussian entangled states except for a small portion of the BS number states. We also show that the entanglement of this small portion is detectable by the measurements of further higher order moments based on the idea of Ref. \cite{Sun11}. 
\subsection{SU(2) and SU(1,1) commutation relations and separable conditions}

Let us define the set of operators
\begin{eqnarray}
J_x &=& \frac{1}{2  }( a ^\dagger b + ab^\dagger), \ \  K_x = \frac{1}{2}( a ^\dagger b^\dagger + ab ), \nonumber \\
J_y &=& \frac{1}{2i }( a ^\dagger b -  ab^\dagger), \ \  K_y = \frac{1}{2i  }( a ^\dagger b^\dagger -  ab ), \nonumber \\
J_z &=& \frac{1}{2}(  N_a -  N_b), \ \  K_z = \frac{1}{2}(  N_a +   N_b+1 ), \label{def211}
\end{eqnarray}
where $ N_a := a ^\dagger a  $ and  $ N_b := b ^\dagger b   $ are number operators of  the local modes. The operators $J_i$ with $i =x,y,z$ form the SU(2) algebra while the operators $K_i$ with $i =x,y,z$ form the SU(1,1) algebra \cite{SV3,HZ06,Aga05,Nha07,Sun11}.  
The commutation relation $[K_y,K_z]= iK_x$ implies the uncertainty relation $\ave{\Delta^2 K_y } \ave{\Delta^2 K_z } \ge \frac{1}{4} |\ave{ K_x }|^2 $. From its partial transposition  we have the separable condition in Eq. (10) of Ref. \cite{Sun11}
\begin{eqnarray}
\left[ \ave{\Delta^2 (J_y)  }+ \frac{1}{4} \right] \ave{\Delta^2 K_z } \ge \frac{1}{4} |\ave{ J_x }| ^2 ,  \label{10Sun11}
\end{eqnarray} where the partial transposition of the second system with respect to the number basis
is calculated by the relation  $\langle  a^{\dagger k} a^l   b^{\dagger o} b^p     \rangle_{PT}=\langle  a^{\dagger k} a^l   (   b^{\dagger o} b^p      )^\dagger   \rangle =\langle  a^{\dagger k} a^l    b^{\dagger p} b^o   \rangle $.
 Similarly, starting from the uncertainty relation $\ave{\Delta^2 J_y } \ave{\Delta^2 J_z } \ge \frac{1}{4} |\ave{ J_x }|^2 $ 
  we have another separable condition
\begin{eqnarray}\left[ \ave{\Delta^2 (K_y)  }- \frac{1}{4} \right] \ave{\Delta^2 J_z } \ge \frac{1}{4} |\ave{ K_x }| ^2 .  \label{m10Sun11}\end{eqnarray}

In the next two subsections we investigate the role of these separable conditions  on the detection of the non-Gaussian entanglement. The entanglement of the present non-Gaussian states is induced by the unitary operators $V_\xi= e^{\xi a^\dagger b^\dagger - \xi^* a b}$  in Eq. (\ref{defA}) and $U_r =e^{r a^\dagger b - r^* a b^\dagger }$  in Eq. (\ref{defC}), and  the Hamiltonians can be connected with a partial transposition as it maps $r a^\dagger b - r^* a b^\dagger \to r a^\dagger b^\dagger - r^* a b $. 
Hence, it might be insightful to find general relationships between the non-Gaussian entanglement and the separable conditions with the terms in the Hamiltonian.

\subsection{Entanglement verification of the TMS number states}
The TMS number states of Eq. (\ref{basisi}) are simultaneous eigenstates of the number operators $\hat A_\xi^\dagger \hat A_\xi  $ and $\hat B_\xi^\dagger \hat B_\xi $. 
 From  the definition of $J_z$ in Eq. (\ref{def211}) and  the relations in 
  Eq. (\ref{defA}) 
 we have  \begin{eqnarray}  \hat A_\xi^\dagger \hat A_\xi  -  \hat B_\xi^\dagger \hat B_\xi   &=&N_a -  N_ b = 2 J_z.  
 \end{eqnarray}
 This implies that the TMS number states are eigenstates of $J_z$ and 
  $\ave{\Delta^2 J_z} =0$. 
In turn, 
 we can write \begin{eqnarray}
K_x &=& \frac{1}{2  (1- |\xi |^2)} [ (  \hat A_\xi^\dagger +\xi^*  \hat B_\xi  )( \hat B_\xi^\dagger  +  \xi^*  \hat A_\xi )  \nonumber \\  & &+    (  \hat A_\xi +\xi  \hat B_\xi^\dagger   )( \hat B_\xi +\xi   \hat A_\xi^\dagger )].
\end{eqnarray} Then, its  expectation value  for the TMS number state $\ket{M,N; \xi}$ of Eq. (\ref{basisi}) is calculated  to be
 \begin{eqnarray}
 \ave{K_x}   &=&\frac{  \xi +\xi^*   }{2(1-  |\xi |^2 )}  \ave{   \hat A_\xi^\dagger   \hat A_\xi   +   \hat B_\xi^\dagger  \hat B_\xi  +1  } \nonumber \\
&=&\frac{  \xi +\xi^*  }{2(1-  |\xi |^2 )} (M+N+1).
\end{eqnarray} This implies $| \ave{K_x}  | >0 $.
Consequently, for the TMS number states of Eq. (\ref{basisi}), the left-hand side    of Eq. (\ref{m10Sun11}) is zero and right-hand side of Eq. (\ref{m10Sun11}) is positive. Hence, we can confirm the inseparability of the TMS number states from the violation of Eq. (\ref{m10Sun11}).

Since the separable condition  of  Eq. (\ref{m10Sun11}) is violated by the TMS number state $\ket{M,N; \xi}$ for any integers $M$, $N$, and $ | \xi | >0 $, it might be valuable to emphasize the mechanism of the entanglement detection. 
 On one hand, the TMS number states exhibit strong classical correlation on the photon number as we have $\ave{\Delta ^2 J_z} =0$. On the other hand, they show coherence between  the (Schmidt) bases $\{\ket{k,k}, \ket{k+1,k+1}\}$ as the pair creation and annihilation of the photons occur coherently. This keeps the term $\ave{ K_x}$ non-zero. Therefore, we can say that the strong classical correlation and the coherence on the Schmidt bases signify the existence of entanglement.

Note that there is a different entanglement criterion that can verify the entanglement of the TMS number states based on the measurement of quadrature moments to the fourth order \cite{namiki10}. This criterion is formulated to  verify the entanglement of the TMS number state when the parameter $\xi$ is specified.  
Note also that test of a sequence of high order separable conditions for TMS number states have been reported in Ref. \cite{Dell06}.

\subsection{Entanglement verification of the BS number states}
The BS number states of Eq. (\ref{p-basisi}) are simultaneous eigenstates of the number operators $c^\dagger c $ and $d^\dagger d $.  From   the relations in Eq. (\ref{defC}) and the definition of $K_z$ in Eq. (\ref{def211}) 
 we have
   \begin{eqnarray} c^\dagger c +d ^\dagger d  &=& a^\dagger a +b ^\dagger b =N_a + N_ b =  K_z -1  . 
       \end{eqnarray}
Hence, the BS number states are   eigenstates of  $K_z$, and we have $ \ave{ \Delta^2 K_z} =0 $.  In turn, from Eq. (\ref{eq18}) and  definition of $J_x$  we have  $ \ave{ J_x} = \frac{ r + r^* }{2(1+|r|^2)} (n-m)$ for the BS number state $\ket{n,m; r }$ of Eq. (\ref{p-basisi}). This implies that the left-hand side of Eq. (\ref{10Sun11}) is zero and the right-hand side of Eq. (\ref{10Sun11}) is positive whenever $n \neq m$. Hence, we can confirm the inseparability of the BS number state from the violation of Eq. (\ref{10Sun11}) except for the case of $n =m $.

To cover the case of $n=m$, we consider another separable condition proposed in Ref. \cite{Sun11}.  Following Eq. (11) of \cite{Sun11}, let us define 
 \begin{eqnarray}
H_x &=& \frac{1}{2  } [( a^\dagger b^\dagger ) ^2  +( a b )^2 ], \nonumber \\
H_y &=& \frac{1}{2 i  } [( a^\dagger b^\dagger)^2  - (  a b  )^2], \nonumber \\
N_+ &=& \frac{1}{4} (N_a +N_b), \nonumber \\
\tilde L_x &=& \frac{1}{2 } [( a^\dagger b)^2  +( a b^\dagger)^2 ], \nonumber \\
\tilde  L_y &=& \frac{1}{2i  } [(  a^\dagger b)^2  -( a b^\dagger)^2 ]. 
 \end{eqnarray}
 From the commutation relation $[H_y, N_+] = iH_x$ we have the uncertainty relation $\ave{\Delta ^2 H_y} \ave{\Delta^2  N_+} \ge \frac{1}{4}|\ave{H_x} |^2 $. Partial transposition on this inequality leads to the separable condition in  Eq. (14) of  \cite{Sun11}, 
\begin{eqnarray}
\left[ \ave{\Delta^2 (\tilde  L_y)  }+ \ave{N_{22} } \right] \ave{\Delta^2 N_{+} } \ge \frac{1}{4} |\ave{ \tilde  L _x }| ^2,  \label{14Sun11}
\end{eqnarray}
where $N_{22}:= \frac{1}{4}[a^2, (a^\dagger )^2 ]\otimes [b^2, (b^\dagger )^2 ] 
 = (2a^\dagger a +1 ) (2b^\dagger b +1) \ge 1$ is a positive operator. 

For the BS number states of Eq. (\ref{p-basisi}) we have $\ave{\Delta ^2 N_+} = \ave{\Delta^2  K_z }=0$, and the left-hand side of Eq. (\ref{14Sun11}) is zero.   
In turn,  by picking the terms which include the same number of annihilation and creation operators from the expression
\begin{eqnarray}
\tilde  L_x
 &=& \frac{1}{2   (1+|r| ^2) ^2  } \{  [(c^\dagger) ^2  +(r^* {d ^\dagger})^2  +2  r^*  c^\dagger d ^\dagger ] \nonumber \\ &&  \times [ d  ^2   +  (r^* c  )^2   - 2{r^*}  c d  ]  \nonumber \\
  & & +  [(d^\dagger) ^2  +(r c ^\dagger) ^2 - 2 r  c^\dagger d ^\dagger ]   [(c ^2) +r ^2 d^2 +2 r c d ]  \}\nonumber \\ 
\end{eqnarray}
  we  have 
\begin{eqnarray}
\ave{\tilde  L_x}&=& \frac{ r^2 +(r^*)^2}{2(1+ |r|^2 )^2} \ave{( c^\dagger)^2 c^2 +(d^\dagger)^2 d^2 -2 c^\dagger c d^\dagger d } 
\nonumber \\
&= &  \frac{  r^2 +(r^*)^2}{2(1+ |r|^2)^2} [m(m-1)+n(n-1)-4 nm ]\nonumber . \\ 
\end{eqnarray}
This implies 
 $|\ave{\tilde  L_x}| =     \frac{   |r^2 +(r^*)^2 |}{(1+ |r|^2)^2 } n(n+1)  > 0$  for the case of $n=m$ .
Hence, the left-hand side of Eq. (\ref{14Sun11}) is zero and right-hand side of Eq. (\ref{14Sun11}) is non-zero for the BS number states with $n=m \neq 0$. Therefore, the violation of the condition of Eq. (\ref{14Sun11}) can verify the  entanglement of the BS number states with $n=m$.

 Similar to the situation that the TMS number states violates the condition of Eq. (\ref{m10Sun11}), the mechanism of entanglement detection of the BS number states can be explained by the coexistence of strong classical correlation and coherence  on the photon-number basis. The classical correlation is confirmed from $\ave{\Delta^2  K_z}=0 $ 
or $\ave{\Delta^2  N_+}=0 $  while the non-zero coherence of $\ave{J_x} >0$ or $\ave{\tilde L_x} >0$ is observed on the (Schmidt) bases with single-photon exchange $\{\ket{n+m-k,k}, \ket{n+m-k-1,k+1} \}$ or the (Schmidt) bases with  two-photon exchange $\{\ket{n+m-k,k}, \ket{n+m-k-2,k+2} \}$ for  $n\neq m $ and $n=m $, respectively.  
 Therefore, we can observe basically the same  structure in  
  non-Gaussian entanglement for the TMS number states and the BS number states.  %

 \begin{figure}[phtb]
  \includegraphics  [width=0.95\linewidth]{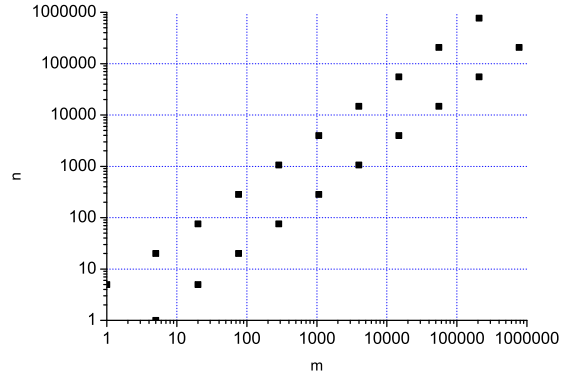}
  \caption{Distribution of $(m,n)$ that satisfies $(n-m)^2 -(n+m)=0$. 
  There are eleven  solutions for $0 \le m < n \le 10^6 $:  
  (0,1), (1,5),
 (5,   20),
(20,    76),
(76,    285),
(285,   1065),
(1065,    3976),
(3976,    14840),
(4840,    55385), 
(55385,    206701), and
 (206701,    771420). 
 The solutions  $(m,n)=\{(0,1),(1,0)\}$ are not displayed  in this figure. 
   }%
    \label{fig:sep-fig3.eps}
\end{figure}

 It might be worth to mention the existence of the BS number states that do not violate the higher order separable condition of Eq. (\ref{14Sun11}).
  The  set of such BS number states is specified by the condition $\ave{\Delta ^2 N_+} =0$,  namely, it can be specified by  the pairs of non-negative integers $(m,n)$ that fulfill   $(n-m)^2 -(n+m)=0$. The distribution of the pairs $(m,n)$  for $1 \le n, m \le 10^6$ is shown in FIG. \ref{fig:sep-fig3.eps}.  Hence, the separable  conditions  of Eq. (\ref{14Sun11})  is also widely useful to verify the entanglement of the BS number states but there is a small portion of the states which fulfill the condition. Consequently, at the present, we require two separable conditions in order to verify whole class of the BS number states based on the measurement of quadrature moments. It is in a sharp contrast to the case of the TMS number states, in which the entanglement can be verified by the violation of the single separable condition of Eq. (\ref{m10Sun11}). 

\section{Summary and remarks}
We have considered two photonic families of entangled states generated from the products of number states by using two-mode Gaussian entangling operations and investigated their inseparability associated with  the entanglement criterion based on the measurements of quadrature moments. 
We have specified the parameter regime of the states where the entanglement cannot be verified by the measurement of the covariance matrix, which covers the case of the measurement of quadrature moments to the second order. We have also shown a couple of continuous-variable entanglement criteria to verify whole the  non-Gaussian entanglement of the two families. An interesting finding is that  the measurement of the covariance matrices is useless to detect substantial part of the non-Gaussian entanglement. On the other hand, it has been shown that the measurements to the fourth order moments are sufficient for the entanglement verification except for a very small portion of the non-Gaussian  states.

  Generation of non-Gaussian entangled states from the number states and Gaussian operation would  be an important goal of experiments and systematic analysis of such states might be an insightful step to comprehend quantum correlation beyond the Gaussian framework.

\acknowledgements
This work was supported by the Grant-in-Aid for the Global COE Program ``The Next Generation of Physics, Spun from Universality and Emergence'' from the Ministry of Education, Culture, Sports, Science and Technology (MEXT) of Japan. 


\end{document}